\documentclass{article}
\usepackage{spconf,amsmath,graphicx,hyperref}

\title{FilterAugment: An Acoustic Environmental Data Augmentation Method}
\name{Hyeonuk Nam, Seong-Hu Kim, Yong-Hwa Park \thanks{This work was supported by “Human Resources Program in Energy Technology” of the Korea Institute of Energy Technology Evaluation and Planning (KETEP), granted financial resource from the Ministry of Trade, Industry \& Energy, Republic of Korea. (No. 20204030200050)}}

\address{Department of Mechanical Engineering, Korea Advanced Institute of Science and Technology, Korea}
\begin{document}
\ninept
\maketitle
\begin{abstract}
Acoustic environments affect acoustic characteristics of sound to be recognized by physically interacting with sound wave propagation. Thus, training acoustic models for audio and speech tasks requires regularization on various acoustic environments in order to achieve robust performance in real life applications. We propose \textit{FilterAugment}, a data augmentation method for regularization of acoustic models on various acoustic environments. FilterAugment mimics acoustic filters by applying different weights on frequency bands, therefore enables model to extract relevant information from wider frequency region. It is an improved version of frequency masking which masks information on random frequency bands. FilterAugment improved sound event detection (SED) model performance by 6.50\% while frequency masking only improved 2.13\% in terms of polyphonic sound detection score (PSDS). It achieved equal error rate (EER) of 1.22\% when applied to a text-independent speaker verification model, outperforming model used frequency masking with EER of 1.26\%. Prototype of FilterAugment was applied in our participation in DCASE 2021 challenge task 4, and played a major role in achieving the 3rd rank.
\end{abstract}
\vspace{-10pt}
\section{Introduction}
\label{sec:intro}
Training deep neural networks (DNNs) requires enormous high-quality dataset in order to regularize models and achieve robust performances. However, collecting such dataset requires tremendous time and cost. Many data augmentation methods have been proposed in order to effectively utilize limited size of dataset. Data augmentation methods increase dataset size by providing various “views” of the same data \cite{autoaug, da4aer, da4esc}. By training neural networks with data from different views every epoch, they can be regularized to learn information shared across different views.

Application of deep learning (DL) methods in audio and speech tasks such as sound event detection (SED), speaker recognition and automatic speech recognition (ASR) have been adopting techniques from DL methods developed for computer vision domain \cite{da4aer, da4esc, asrreview, cnn4audioclassification, crnn4sed}. Applying short-time Fourier transform (STFT) \cite{dsp} on audio data can change data dimension from 1D waveform (time) to 2D spectrogram (time and frequency), which can be treated like image data \cite{cnn4audioclassification, crnn4sed, coughcam}. Some of data augmentation methods proposed for computer vision tasks such as mixup \cite{mixup} are actively adopted in audio and speech domain. However, most of the image data augmentation methods including rotation, flip, shear and crop \cite{autoaug} result in irrelevant transform of audio data when applied on spectrograms. Therefore, data augmentation methods consistent with acoustics and signal processing domain knowledge are required to effectively train \textit{acoustic models} in audio and speech domain. The term acoustic model usually refers to the encoder structures in ASR models for extracting acoustic information from speech audio signals. In this paper, we call encoder structures in neural networks for audio and speech tasks as acoustic models, as their common purpose is to effectively extract useful acoustic information from audio signals.

Early works introducing DL methods on audio and speech tasks used conventional audio signal processing methods \cite{da4aer, da4esc} for data augmentation. Although conventional audio signal processing methods do help increasing dataset size, they are not easy to utilize without sufficient understanding in acoustics and signal processing domain. This problem was resolved by SpecAugment \cite{specaug}, involving time masking and frequency masking those simply mask a small time and frequency range of mel spectrograms. Time and frequency masking can be easily applied to train acoustic models as their algorithms are simple and straightforward, but they are brutal in the sense that they completely remove certain information from the data.

In this work, we propose \textit{FilterAugment}, an improved version of frequency masking from SpecAugment \cite{specaug}. FilterAugment is proposed to regularize acoustic models over various acoustic environments by mimicking acoustic filters. Sound could be heard differently in various acoustic environments such as conference room, shower room, performance hall, cave, etc. Although such acoustic characteristics derived from different physical contexts vary a lot, human can recognize sound events, speakers, or spoken words regardless of acoustic characteristics. These highly variable acoustic characteristics can be modeled using acoustic filters \cite{acoustics}, which FilterAugment aims to mimic in a simplified way so that acoustic models could be trained to recognize the sound contents in various acoustic environments. FilterAugment approximates acoustic filters by applying random weights on randomly determined frequency bands. Although applying FilterAugment does not make resulting to sound as natural as results of applying acoustic filters, it effectively regularizes acoustic models by extracting sound information from wider range of frequency. Prototype of FilterAugment was applied on our participation in Detection and Classification of Acoustic Scenes and Events (DCASE) 2021 challenge task4, resulting in the 3rd rank \cite{mytechreport}. Considering that most of other teams above 5th rank added major modifications on model architecture \cite{dcase1, dcase2, dcase3, dcase4, dcase5} while we did not, FilterAugment is proven to be a powerful data augmentation method. The official implementation code for FilterAugment applied on SED is shared on GitHub\footnote{https://github.com/frednam93/FilterAugSED/}.

\vspace{-5pt}
\section{Audio Data Augmentations}
\label{sec:audio data augmentations}
Data augmentation methods in audio and speech domain includes conventional audio signal processing methods such as time stretching, pitch shift, clipping, suppressing, adding noise, adding reverberation, etc. \cite{da4aer, da4esc}. These methods reflect domain knowledge in acoustics and signal processing, thus they have been frequently adopted for data augmentation purpose. However, data augmentation using conventional audio signal processing methods could introduce some inefficiencies when training acoustic models. Applying conventional audio signal processing methods requires prior knowledge to appropriately handle audio data. In addition, these methods may involve more computations in expense for more natural sound, which does not even guarantee to train acoustic models better. Such inefficiencies hinder optimal training of acoustic models. Therefore, we need data augmentation methods that are simple, intuitive, yet effective for training acoustic models to learn to extract information from audio data.

SpecAugment \cite{specaug} is one of the most powerful and widely used data augmentation methods in audio and speech domain. Instead of applying data augmentation on waveform, it proposed time warping, time masking, and frequency masking those could be directly applied on log mel spectrogram. As it is applied directly on the input feature space, it is easy to comprehend and use. Intuitively, applying time warping on audio would sound like the audio played faster in some points and slower in some other points. Time masking would sound that some parts are not played for short duration. Frequency masking would sound like some part of frequency range is missing. As long as these distortions are not too severe, human can recognize the content of audio data after these processing, and trained acoustic models should do as well. Although these methods do not sound as natural as conventional audio processing methods when transformed back in waveform, it helps training acoustic models more effectively with extreme cases.

\begin{figure*}[ht]
  \centerline{\includegraphics[width=17.5cm]{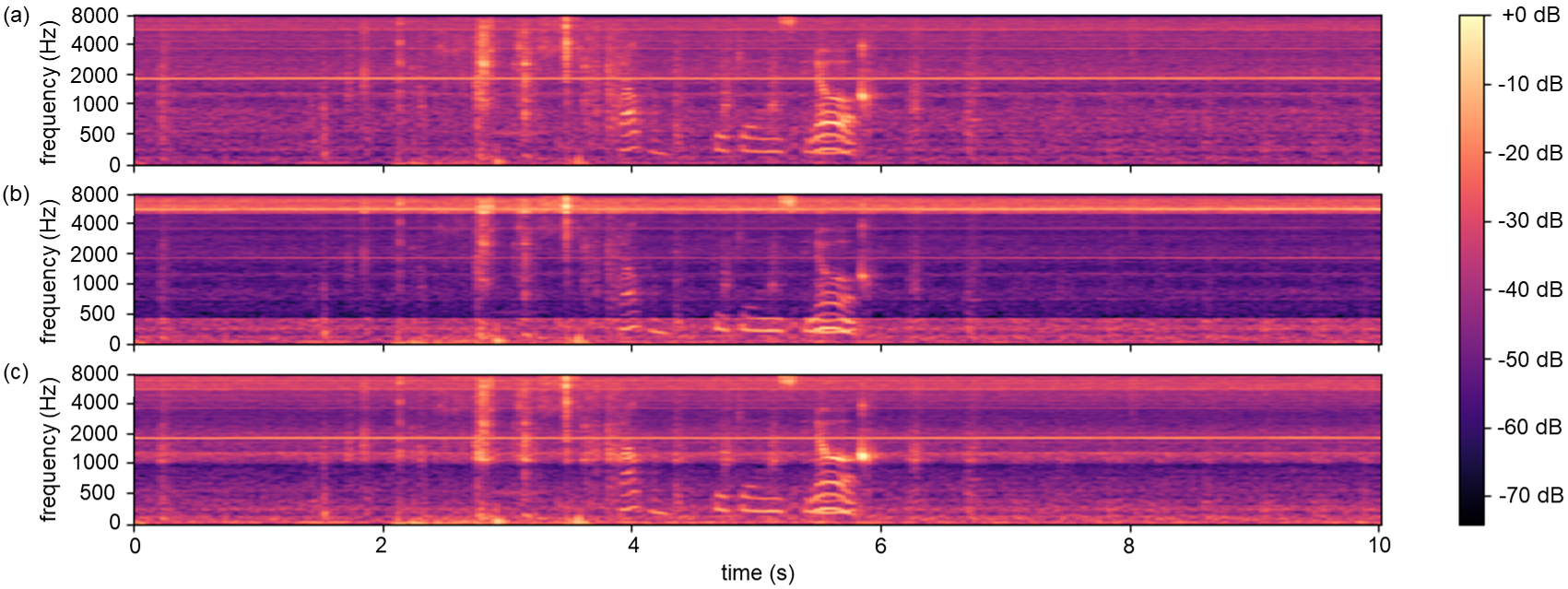}}
\vspace{-10pt}
\caption{Illustrations on application of FilterAugment on an example audio clip. (a) is log mel spectrogram of the original audio clip, (b) is log mel spectrogram applied with step type FilterAugment, (c) is log mel spectrogram applied with linear type FilterAugment.}
\label{fig:mels}
\vspace{-15pt}
\end{figure*}

\vspace{-10pt}
\section{Proposed Method}
\label{sec:proposed method}
\subsection{Motivation}
FilterAugment can be explained in two different but related points of view. In the viewpoint of acoustics and signal processing, FilterAugment regularizes acoustic models to various acoustic environments by mimicking acoustic filters. From the viewpoint of acoustic model training, FilterAugment learns to effectively extract acoustic information from wide frequency ranges while training. We will first explain motivation of FilterAugment in terms of acoustics and signal processing viewpoint, then discuss its significance in terms of acoustic model training.

When we hear sound events or speeches, we can recognize their contents regardless of acoustic environments unless the environments are too noisy or too echoic. It is because our auditory system is trained to understand the sound contents regardless of the acoustic environments. Acoustic environment refers to the physical objects surrounding the sound source, receiver (ear or microphone) and the air surrounding them (medium of the sound wave propagation). These interact with sound wave and change the acoustic characteristics of sound perceived by receiver with absorption, reflection, scattering, etc. \cite{acoustics}. Such change in acoustic characteristics appears as relative change in energy on different frequency range. For example, when the sound source is far away from the receiver, high-frequency energy reduces as it dampens more than low-frequency energy does while propagating in the air. Similarly, when there is a wall or any object blocking between the receiver and the sound source, high-frequency energy reduces as it does not diffract easily thus does not propagate to the receiver much. In addition, room’s walls and furnishings cause reverberation, and early reverberation causes coloration which alters acoustic characteristics of the sound perceived. Such change in energy on different frequency ranges can be simulated by designing appropriate types of filter: high pass filter, low pass filter, band pass filter, notch filter, etc. \cite{dsp, acoustics}. However, designing and applying such filters for data augmentation purpose requires understanding in acoustics and signal processing. In addition, applying filters to training audio data takes time to compute filters’ impulse responses and convolute them with audio data. Although training time might not increase that much, it will complicate training and optimization process. Therefore, we propose FilterAugment, a simpler alternative data augmentation method to mimic filter effect. FilterAugment randomly increases or decreases energy of random frequency ranges of log mel spectrograms. Such increase or decrease of energy in random frequency range is equivalent to application of random filters. Although it might sound unnatural compared to acoustic filters as it induces discrete filter design, FilterAugment is much easier to comprehend and use.

From the viewpoint of acoustic model training, randomly weighting on random frequency bands of log mel spectrogram enables training of acoustics models to extract sound information from wider frequency regions. Without FilterAugment, acoustic model is likely to learn to recognize frequency ranges that exhibit dominant and distinctive feature of desired labels. However, we can recognize the sound content regardless of the acoustic environment that might even drastically reduce the frequency region with the most distinctive feature. It means that we still can recognize sound content from the other less distinctive frequency ranges. This would be the reason why applying frequency masking \cite{specaug} improves training acoustic models as well. As frequency masking removes information from certain random frequency range, it helps to train acoustic model to infer the sound information from less distinctive frequency regions too. However, frequency masking completely removes certain part of energy that might help inferring the sound information. Such brutal damage on spectrum not only rarely happens in real situations but also causes the model to be trained to forcibly extract information from indistinct frequency ranges. Therefore, FilterAugment weakens some parts of frequency range while strengthening other parts instead. Lowering energy instead of removing it would at least let acoustic models to infer the information from that frequency region. In addition, increasing other frequency range energy would train acoustic models to recognize sound information from various frequency region as they will be trained with the same data highlighted on different frequency region every epoch. Therefore, FilterAugment helps training acoustic models to extract information from the wider range of frequency regardless of each frequency’s relative significance composing the sound information.

\begin{figure}[ht]
\centerline{\includegraphics[width=8.5cm]{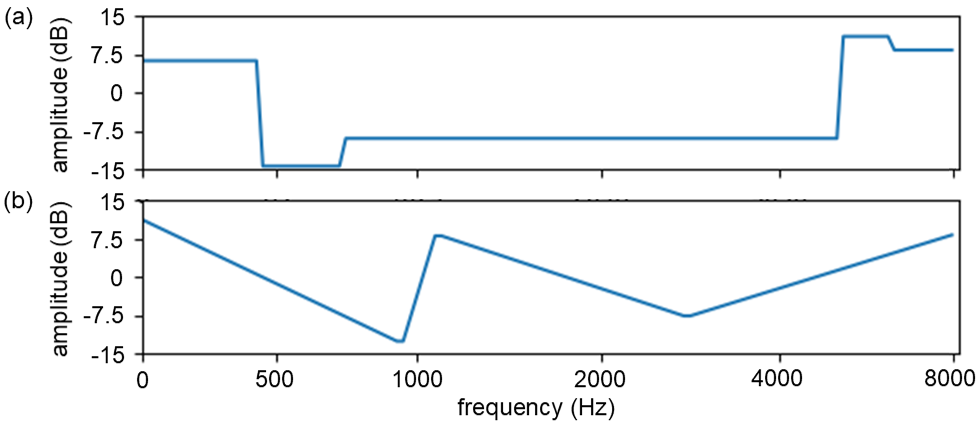}}
\vspace{-10pt}
\caption{Filters applied on the example audio clip. (a) is step type filter that resulted on Fig. \ref{fig:mels}. (b), and (b) is linear type filter that resulted on Fig. \ref{fig:mels}. (c).}
\label{fig:filters}
\vspace{-15pt}
\end{figure}

\subsection{Algorithm}
We propose three types of FilterAugment: step, linear and mixed type. Detailed algorithm for step type FilterAugment is as follows.

\begin{enumerate}
    \itemsep0em
    \item{Randomly choose number of frequency bands ${n}$ within hyperparameter \textit{band number range}.}
    \item{Randomly choose ${n - 1}$ mel frequency bins between 0 and ${F}$ (number of mel frequency bins in mel spectrogram), and include 0 and ${F}$ to form ${n + 1}$ frequency boundaries. These frequency boundaries are separated from each other at least by hyperparameter \textit{minimum bandwidth.}}
    \item{Randomly choose ${n}$ different weights within hyperparameter \textit{dB range}. }
    \item{Add chosen ${n}$ weights on ${n}$ frequency bands of log mel spectrogram defined by each set of subsequent frequency boundaries respectively.}
\end{enumerate}

\noindent
As a result, mel spectrogram’s energy is amplified in some frequency bands while reduced in other bands. Note that \textit{Minimum bandwidth} is adopted to prevent applying weights too locally. Amplifying or reducing energy of a frequency band with too narrow bandwidth would cause negligible change in how audio clips sound, so we set minimum bandwidth to make sure each weight cause significant change in sound. Step type FilterAugment is the simplest type of FilterAugment composed of series of step functions. An example of step type FilterAugment is shown in Fig. \ref{fig:mels}. (b). The filter applied on the original audio clip’s log mel spectrogram Fig. \ref{fig:mels}. (a) to produce it is shown in Fig. \ref{fig:filters}. (a). It can be observed that when compared to the original log mel spectrogram, the augmented result has higher energy below 400 Hz and above 5 kHz while the rest of frequency regions have lower energy. Also, the abrupt change in energy across frequency range can be seen as clear horizontal lines on frequency boundaries at 400 Hz and 5 kHz on Fig. \ref{fig:mels}. (b). 

Abrupt energy changes at boundary frequencies on step type FilterAugment cause unnatural sound. To make more naturally augmented audio data, linear type FilterAugment is proposed. Algorithm for linear type FilterAugment shares 1st and 2nd steps with step type FilterAugment. Rest of the algorithm is as follows.
\begin{enumerate}
    \setcounter{enumi}{2}
    \itemsep0em
    \item{Randomly choose ${n + 1}$ different weights corresponding to ${n + 1}$ frequency boundaries within hyperparameter \textit{dB range}.}
    \item{Linearly interpolate weights within frequency boundaries.}
    \item{Add resultant interpolated weights on log mel spectrogram.}
\end{enumerate}

\noindent
While step type FilterAugment applies discontinuous filter that is composed of series of step functions, linear type applies continuous (although not differentiable) filters that are composed of series of linear functions. An example of linear type FilterAugment is shown in Fig \ref{fig:mels}. (c). It is produced by applying filter Fig. \ref{fig:filters}. (b) to original log mel spectrogram Fig. \ref{fig:mels}. (a). Compared to the original log mel spectrogram, peaks around 0 Hz, 1.2 kHz and 8 kHz and dips around 900 Hz and 3 kHz can be observed and the gradual change between these peaks and dips can be observed as well. Linear type FilterAugment shows smoother change in energy across frequency axis in Fig. \ref{fig:mels}. (c) when compared step type FilterAugment in Fig. \ref{fig:mels}. (b) that shows abrupt change in energy across frequency axis. 

As step and linear type FilterAugment are expected to have different effect on training acoustic models, mixed type FilterAugment is proposed to train acoustic models to be regularized on both step and linear type of FilterAugment. Hyperparameter \textit{mix ratio} determines the probability of using step type FilterAugment. For example, if mix ratio is 0.7, then there is 70\% chance that step type FilterAugment is applied to the batch and 30\% chance that linear type FilterAugment is applied to the batch.

\vspace{-5pt}
\section{Experiments}
\label{sec:experiments}
\subsection{Implementation Details}
FilterAugment algorithm was tested on SED and text-independent speaker verification: one task each from audio and speech domains. We first optimized hyperparameters of FilterAugment and frequency masking on SED, then applied them with the optimized hyperparameters on speaker verification. Frequency masking involves a hyperparameter \textit{maximum masking ratio} which determines the maximum ratio of mel frequency bins to be randomly masked during the training to $F$. The performances of the baseline models with and without FilterAugment and frequency masking are compared.

SED baseline model in this work is an upgraded version of baseline model for DCASE 2021 challenge task 4 \cite{DCASEtask4, dcasewebsite}, which is the same with optimized model in \cite{mytechreport} without the prototype FilterAugment. From DCASE baseline \cite{dcasewebsite, dcasebaseline}, dimensions of convolutional recurrent neural network (CRNN) are doubled. Activation functions in convolutional neural network (CNN) structure are replaced by context gating \cite{contextgating}. Waveforms are normalized so that their absolute maximum equals to one. Time masking \cite{specaug} is added with optimized masking range within 7 -- 30 frames (0.11 -- 0.48 seconds). Weak prediction masking is applied to test predictions \cite{mytechreport}. We compared the performance of baseline model with and without frequency masking and step/linear/mixed type FilterAugment. Evaluation metrics measured include polyphonic sound detection score (PSDS) criteria on DCASE 2021 challenge task 4 \cite{DCASEtask4, dcasewebsite, PSDS} for two situations (PSDS$_{1}$ and PSDS$_{2}$), macro collar-based F1 score \cite{sedmetrics} and macro intersection-based F1 score \cite{insightdcase2020}. PSDS$_{1}$ penalizes more on inaccurate time localization while PSDS$_{2}$ penalizes more on confusions between classes. These four metrics are higher with better SED performance. Hyperparameters for frequency masking and FilterAugment are optimized for the highest PSDS$_{1}$ + PSDS$_{2}$ which is official evaluation score on DCASE 2021 challenge task 4. F1 scores are listed for reference.

We applied the optimized settings of frequency masking and FilterAugment to text-independent speaker verification baseline model, which is the model without data augmentation from \cite{speakerverification}. Then, only dB range was re-optimized because the datasets \cite{voxceleb1, voxceleb2} are composed of interviews from YouTube videos recorded in controlled acoustic environments. Although they might have some noises, speaker-microphone distances are usually close and the microphones’ recording qualities are good enough to keep speeches’ acoustic characteristics almost constant. Therefore, dB range of FilterAugment is narrowed to match the variance of acoustic characteristics of speaker verification task. The baseline model is ResNet-34 with SE module and attentive statics pooling (ASP) \cite{ASP}. It is trained using 5994 speakers of Voxceleb2 dataset \cite{voxceleb2} with combined loss function composed of Angular Prototypical (AP) loss \cite{APloss} and vanilla softmax loss. Speaker embeddings are extracted from each utterance of Voxceleb1 dataset \cite{voxceleb1} and compared using cosine similarities for validation. For evaluation metrics, we used equal error rate (EER) and minimum detection cost function (MinDCF) with C$_{miss}$ = 1, C$_{fa}$ = 1 and P$_{target}$ = 0.05 \cite{voxceleb1, MinDCF}. Lower EER and MinDCF values imply better speaker verification performance.
\subsection{Results and Analysis}

\begin{table}[t]
\caption{Performance of SED models trained with and without frequency masking and FilterAugment.}
\vspace{5pt}
\centering
\begin{tabular}{l|llll}
\hline
\textbf{Methods} & \textbf{PSDS$_{1}$ $\uparrow$} & \textbf{PSDS$_{2}$ $\uparrow$} & \textbf{CB-F1 $\uparrow$} & \textbf{IB-F1 $\uparrow$} \\ \hline
baseline         & 0.387          & 0.598          & 0.477          & 0.708          \\
freq masking     & 0.396          & 0.610          & 0.470          & 0.710          \\
step FiltAug     & 0.412          & 0.634          & 0.474          & 0.712          \\
linear FiltAug   & \textbf{0.413} & \textbf{0.636} & \textbf{0.490} & \textbf{0.735} \\ \hline
\end{tabular}
\vspace{-15pt}
\label{tab:sed}
\end{table}

\begin{table}[t]
\caption{Performance of SED models trained using mixed type FilterAugment. None of these surpasses neither step type nor linear type FilterAugment.}
\vspace{5pt}
\centering
\begin{tabular}{l|llll}
\hline
\textbf{mix ratio} & \textbf{PSDS$_{1}$ $\uparrow$} & \textbf{PSDS$_{2}$ $\uparrow$} & \textbf{CB-F1 $\uparrow$} & \textbf{IB-F1 $\uparrow$} \\ \hline
0.9                & 0.407          & 0.628          & 0.473          & 0.719          \\
0.7                & 0.401          & 0.606          & 0.469          & 0.709          \\
0.5                & 0.395          & 0.602          & 0.476          & 0.713          \\
0.3                & 0.401          & 0.610          & 0.472          & 0.710          \\
0.1                & 0.408          & 0.622          & 0.470          & 0.711          \\ \hline
\end{tabular}
\vspace{-15pt}
\label{tab:mix}
\end{table}

Optimized hyperparameter for frequency masking is \textit{maximum masking ratio} = 1/16. Optimized hyperparameters for step type FilterAugment (listed as step FiltAug in Table \ref{tab:sed}) are \textit{dB range} = (-6, 6), \textit{band number range} = (2, 5), and \textit{minimum bandwidth} = 4. Optimized hyperparameters for linear type FilterAugment (listed as linear FiltAug in Table \ref{tab:sed}) are \textit{dB range} = (-6, 6), \textit{band number range} = (3, 6), and \textit{minimum bandwidth} = 6. Mixed type FilterAugment uses hyperparameters optimized for step and linear type FilterAugment above. Macro collar-based F1 score and macro intersection-based F1 score are listed in Table \ref{tab:sed} as CB-F1 and IB-F1 respectively. The optimized models’ metric values are listed in Table \ref{tab:sed}, and these values chosen to be displayed are the maximum values of each metric upon three independently trained SED models. Since each training results in separate student model and teacher model by mean teacher method \cite{meanteacher}, these results are the maximum values among the results of 6 models. The results show that FilterAugment improved SED model performance and significantly outperformed model trained using frequency masking. Linear type FilterAugment shows slightly better performance than step type does. The difference is not significant, but more realistic simulation of acoustic filter might helped training acoustic models better. Note that optimized linear FilterAugment requires more frequency bands and wider minimum bandwidth. This should be required to give more distortion on the spectrogram, as linear type FilterAugment tend to cause spectrogram less distortion due to the linear interpolation between two weights on frequency boundaries. Mixed type FilterAugment performs worse than both step and linear type FilterAugment, as shown in Table \ref{tab:mix}. It can be observed that as mixing ratio is closer to 0.5 meaning uniform mixing, the performance worsens. It can be concluded that using different data augmentation method on different batchs during training could result in inconsistent training thus degrade performance. In the end, the best score is achieved by linear FilterAugment. Frequency masking surpassed baseline model performance by 2.13\%, while linear type FilterAugment surpassed baseline model performance by 6.50\%.

\begin{table}[t]
\caption{Text-independent speaker verification performances with and without frequency masking and FilterAugment.}
\vspace{5pt}
\centering
\begin{tabular}{l|ll}
\hline
\textbf{Methods}                       & \textbf{EER (\%) $\downarrow$} & \textbf{MinDCF $\downarrow$} \\ \hline
baseline (no data aug) \cite{speakerverification} & 1.29              & 0.091           \\
frequency masking                      & 1.26              & 0.092           \\
FilterAugment                          & \textbf{1.22}     & \textbf{0.088}           \\ \hline
\end{tabular}
\vspace{-15pt}
\label{tab:sv}
\end{table}

We compared the text-independent speaker verification performance with and without data augmentation methods, and the results are shown in Table \ref{tab:sv}. FilterAugment used for speaker verification follows the same setting of linear type FilterAugment for SED, with re-optimized \textit{dB range} = (-1.5, 1.5). It is shown that FilterAugment shows the better performance than the models without augmentation and with frequency masking. Although Voxceleb1 and 2 have constrained acoustic environments, FilterAugment still outperformed frequency masking.

\vspace{-5pt}
\section{Conclusion}
\label{sec:conclusion}
FilterAugment is an audio data augmentation method that enables effective training of acoustic models in audio and speech domain tasks. It regularizes acoustic models over various acoustic environments by learning to extract sound information from wider frequency range. On both SED and text-independent speaker verification, we showed that FilterAugment outperforms not only models without data augmentation, but also models with frequency masking which uses similar approach with FilterAugment. In conclusion, FilterAugment is simple yet one of the most powerful audio data augmentation methods, having contributed largely to winning 3rd rank in DCASE 2021 task 4.

\vspace{-5pt}
\section{Acknowledgements}
\label{secacknowledgements}
We would like to thank Junhyeok Lee from MINDs Lab Inc. and Won-Ho Jung from KAIST for valuable discussions.



\vfill\pagebreak

\bibliographystyle{IEEEbib}
\bibliography{references.bib}

\end{document}